\documentclass[aps,prl,preprint,superscriptaddress]{revtex4}
\usepackage{graphicx}

\begin{document}
\title{Observation of  pinning mode in    Wigner solid  of $1/3$ fractional quantum Hall excitations  }
\author{Han Zhu}
\affiliation{Princeton University, Princeton, NJ 08544, USA}
\affiliation{National High Magnetic Field Laboratory, Tallahassee, FL 32310, USA}

\author{Yong P. Chen$^*$}
\affiliation{Purdue University, West Lafayette IN, USA}

\author{P. Jiang\footnote{These authors  contributed equally to this paper.}  }
\affiliation{National High Magnetic Field Laboratory, Tallahassee, FL 32310, USA}
 \affiliation{Princeton University, Princeton, NJ 08544, USA}

\author{L. W. Engel}
\affiliation{National High Magnetic Field Laboratory, Tallahassee, FL 32310, USA}

\author{D. C. Tsui}
\affiliation{Princeton University, Princeton, NJ 08544, USA}

\author{L. N. Pfeiffer}
\affiliation{Princeton University, Princeton, NJ 08544, USA}

\author{K. W. West}
\affiliation{Princeton University, Princeton, NJ 08544, USA}
\date{\today}

\begin{abstract}
We report the observation of a resonance  in the microwave spectra of the real diagonal conductivities of  a two-dimensional electron system  within a range of $\sim\pm0.015$ from filling factor $\nu=1/3$.   The resonance  is remarkably   similar to resonances  previously observed near integer $\nu$, and  is interpreted as the collective pinning mode  of  a disorder-pinned Wigner solid phase  of $e/3$-charged carriers .
\end{abstract}

\maketitle

For two-dimensional electron systems (2DES) of low disorder  in high magnetic field ($B$),  interactions between carriers have astounding consequences.   Most striking of these is the fractional quantum Hall effect  (FQHE ) \cite{fqhorig},  which is  the signature of   incompressible liquid states \cite{laughlinorig} at rational fractional Landau level filling.    The  excitations  of the fractional quantum Hall liquid    
carry  fractions of the electronic charge \cite{laughlinorig} as verified  in recent experiments \cite{goldman,shot,smet}.  Analogous to electrons, these fractionally charged excitations themselves may be localized by a disorder potential, or  can   interact to form other states, most notably    other fractional quantum Hall liquids \cite{panfqhcf}.   This paper will present evidence for  an entirely different interacting state of such quasiparticles: a  Wigner solid.  Owing to inevitable pinning of the crystal by disorder, this state is insulating as well.

 Wigner solids occur in   2DES for  small Landau level fillings ($\nu$) \cite{lozoyudson,lamgirvin,kunwc},  when the kinetic energy of  carriers is   quenched by the magnetic field, so that the carriers become effectively dilute, with their
 typical spacing larger than the magnetic length $l_B=(\hbar/eB)^{1/2}$.   At least for 2DES samples of extremely low disorder,  Wigner  solids can  occur as well within the ranges of integer quantum Hall effect plateaus \cite{iqhwc}, in which there is a dilute population  of electrons or  holes in a partially filled Landau level, along with one or more completely filled Landau levels.      Within the IQHE,  Wigner solids are composed of carriers generated
  by  moving $\nu$ away from its integer value,  the insulating property of the pinned solid gives the quantized Hall plateau, which is concomitant with vanishing diagonal conductivity, its finite width.

  The pinning of the Wigner solid by residual disorder \cite{msreview}, besides causing them to be insulators and producing    finite correlation lengths for crystalline order,  gives rise to  a striking rf or microwave resonance in the spectrum \cite{SSCReview,lidensity,yewc,yongab,iqhwc,Zhu}.  This resonance is understood as a pinning mode, or 
 a collective  oscillation of solid domains   within the  disorder potential.       Pinning modes \cite{SSCReview}  can be seen  both  in the low $\nu$ state \cite{lidensity,yewc,yongab},  that  terminates the FQHE series, and within   IQHE's of samples with extremely low disorder \cite{iqhwc,Zhu}. 


    
   This paper reports the observation of a microwave resonance  within the $\nu$ range of  the 1/3 FQHE.  The resonance is  interpreted as pinning mode  of  a Wigner solid of 1/3 charged excitations, and this interpretation is supported by a comparison of    the parameters of the resonance  near $1/3$    with those of  a pinning resonance observed near $\nu= 1$ in the same sample. Both the resonance near $\nu=1/3$ and the resonance near $\nu=1$ show an increase of  peak frequency, $f_{pk}$, as quasiparticle charge density decreases,  and   the integrated absorptions of both resonances suggest  that much of the quasiparticle charge density is participating in the mode.      The  resonance around $1/3$ 
 occurs for fillings within  $\pm 0.015$ of 1/3, much narrower than the $\pm 0.16$ range  of resonance within the IQHE around $\nu=1$, and for the same quasiparticle number density, $\tilde{n} $, the resonance peak frequencies around 1 and 1/3 are nearly the same.


    



The data presented are  from a 2DES in a  50 nm wide GaAs/AlGaAs/GaAs quantum well (\#7-20-99.1)   with  electron density $n=1.1\times10^{11}/\text{cm}^2$, and  low temperature mobility $\mu=15\times10^6\ \text{cm}^2/\text{Vs}$. The microwave spectroscopy technique is similar to those reported earlier \cite{lidensity,yewc,yongab,iqhwc,Zhu,SSCReview}.  As illustrated in Fig. 1(a), we deposited on the sample surface a meandering metal-film coplanar waveguide, with propagation length $l=28$ mm and a slot of width $W=30\ \rm{\mu m}$  separating a driven center conductor from broad, grounded,  side planes.     The microwaves couple capacitively to the 2DES underneath. 
We normalize the transmitted signal, $P$,  taken at a given $\nu$ to that at  a reference integer or rational fractional   filling factor.    In this paper we choose this reference $\nu$ as $\nu_c$ for the particular QHE  under consideration, to minimize   effects of 1) drift  as the superconducting magnet is swept, due to the variation of liquid He levels in the cryostat  and to fluctuation in   the    temperature  of the room, and  2)   background $B$-dependence not due to the 2DES of the  transmitted signal.  Such background $B$ dependence originates from magnetoresistance of the transmission line metal film, and from impurities in the GaAs substrate. 
    From the measured  $P$, we  calculate changes in the real part of the 2DES diagonal conductivity from that at  $\nu_c$,     the quantizing $\nu$ of the QHE being studied,  as    $\text{Re} [\Delta\sigma_{\text{xx}}(\nu)-\sigma_{\text{xx}}(\nu_{c}) (f)]= (W/2lZ_0)\text{ln}P(\nu)/P(\nu_{c})$.   This $Z_0=50 \ \Omega$ is the characteristic impedance of  the coplanar waveguide in the limit of zero 2DES conductivity.  The sample temperature was at 40 mK for data in Fig. 1 and 50 mK for data in Fig. 2.

\begin{figure}
		\includegraphics[width=.48\textwidth]{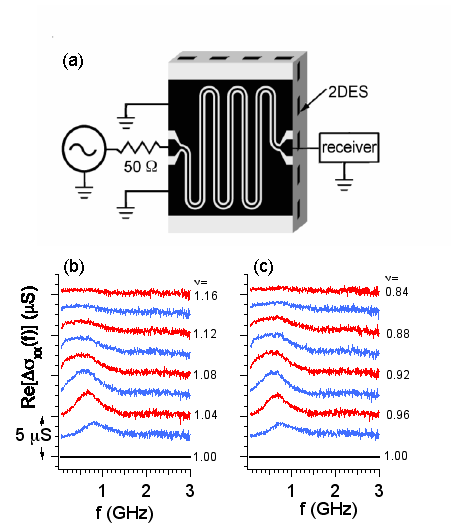}
\caption{\label{Fig1} (color online). (a) Schematic of the   microwave measuring setup, and sample with coplanar waveguide transmission line, with metal film shown as black. b) Real part of change in  diagonal conductivity from that at Landau filling $\nu=1$,   $\text{Re}\left[\Delta \sigma_{\text{xx}}\left(f\right)\right]$,    vs frequency,  $f$, for  $\nu$ increasing from $1$ (bottom trace) to $1.16$ (top trace), in equal steps of $0.02$. Successive traces are offset by $2\ \mu S$ from each other for clarity.  c) $\text{Re}\left[\Delta \sigma_{\text{xx}}\left(f\right)\right]$,    vs  $f$, for  $\nu$ decreasing  from $1$ (bottom trace) to $0.84$ (top trace), in  equal steps of $0.02$. Successive traces are offset by $2\ \mu S$ from each other.  }
\end{figure}

\begin{figure}
		\includegraphics[width=.465\textwidth]{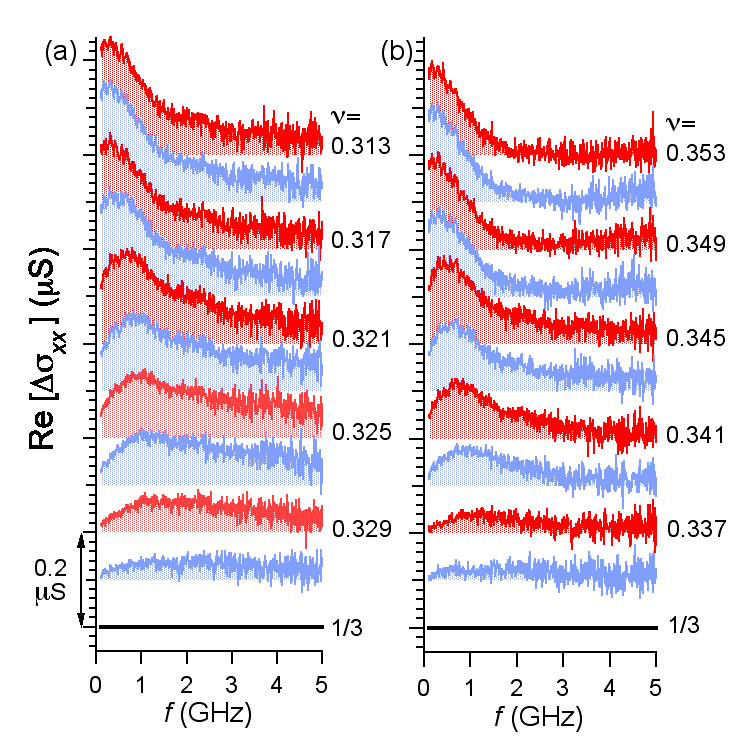}
\caption{\label{Fig2} (color online).  $\text{Re}\left[\Delta\sigma_{\text{xx}}\left(f\right)\right]$,  here the change in real part of the diagonal conductivity  from that at Landau filling $\nu=1/3$, vs frequency $f$  Successive  traces were taken at intervals of 0.002 in $\nu$, and are offset by $0.1\ \mu S$ from each other for clarity.   (a) $\nu$   from $1/3$ (bottom trace),   to $0.313$ (top trace)  (b) $\nu$   from $1/3$ (bottom trace), to $0.353$ (top trace).   }
\end{figure}

We will explicitly compare the  resonances found near $\nu=1/3$ with those  near $\nu=1$, measured  in the same sample.   Fig. 1b shows a series of spectra,  $\text{Re}\left[\Delta\sigma_{\text{xx}}  \right]$  vs $f$ for  $\nu$ between $\nu=1.02$ and $1.16$ in steps of 0.02.
 As in ref. \cite{iqhwc}, the spectra near $\nu=1$ show clear resonances that we ascribe to pinning modes of the IQHWC. 
As  $\nu$ increases  from 1  the peak frequency decreases and the intensity of the resonance  increases; as  $\nu$ increases above 1.16  , the resonance weakens and eventually vanishes.  Spectra taken for $\nu<1$, likewise in steps of $0.02$, are shown in Fig 1c,  and  demonstrate that  the resonance evolves symmetrically about  $\nu=1$, again vanishing for $\nu$ below about 0.84. 
%
%
%
 
Fig 2 presents the  main results of this paper, the spectra of the  resonance within the range of the 1/3 FQHE.  Fig. 2(a) shows   traces of $\text{Re}\left[\Delta\sigma_{\text{xx}} \right]$ vs $f$  for   $\nu$ decreasing from 0.331 to 0.319, and   Fig. 2(b) shows traces for $\nu$ increasing from 0.335 to 0.353.    For both panels   of the figure, $\nu_{ref}=1/3$, and the spectra are taken at $\nu$ separated by steps of only 0.002.    To facilitate the discussion, we denote $\tilde{\nu}=\nu-\nu_c$, where $\nu_c=1$ or $1/3$ is the quantizing filling of the quantum Hall effect.   The dependence on  $\tilde{\nu}$ of the resonance near $\nu=1/3$   strongly resembles that seen near $\nu=1$ in Fig. 1b and c:  as $|\tilde{\nu}|$ increases from zero the   resonance near 1/3  initially increases in intensity,  
 and decreases in peak frequency.  As $|\tilde{\nu}|$ increases  further,  to around $|\tilde{\nu}|\sim 0.015$, the resonant peak in $\text{Re}\left[\Delta\sigma_{\text{xx}}\left(f\right)\right]$ vs $f$  disappears, and the resonance near 1/3 merges into a monotonically decreasing curve.     Different from the spectra near $\nu=1$, the  range of $\nu$ around 1/3 for which a resonance is observed is much narrower, and the peak conductivities of the spectra are much smaller, necessitating the  above-described steps   to suppress  instrumental drift and background absorption.  


\begin{figure}
		\includegraphics[width=.48\textwidth]{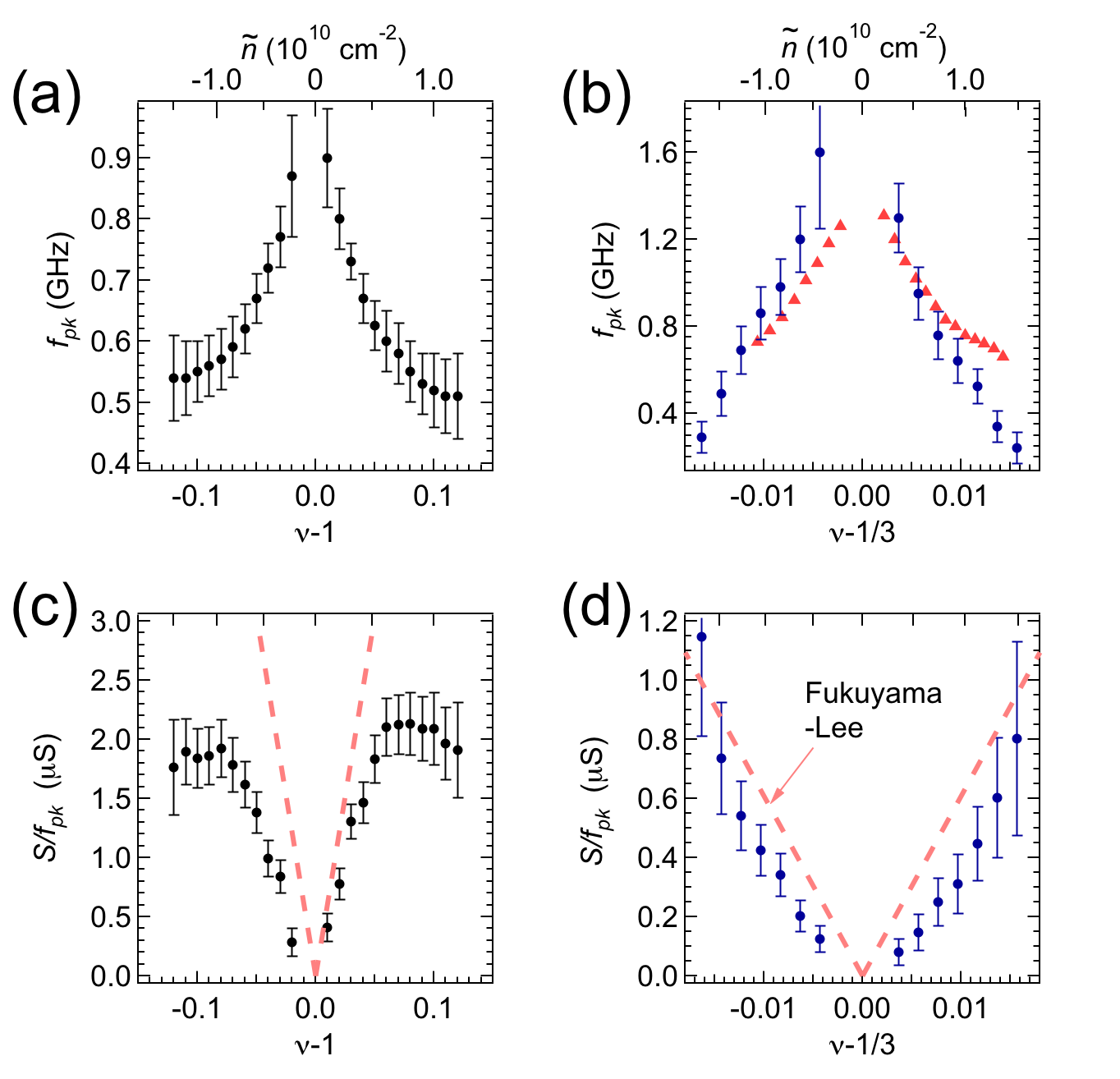}
\caption{\label{Fig3} (color online). a) The resonance peak frequencies $f_{\rm{pk}}$ near filling factor $\nu=1$,  vs   $\nu$ (lower axis) and quasiparticle/hole density, $\tilde{n}$ (upper axis).  b)   filled circles:  $f_{\rm{pk}}$ near $\nu=1/3$ vs $\nu$ (lower axis) and   $\tilde{n}$ (upper axis);  triangles: $f_{\rm{pk}}$  vs $\tilde{n}$ on upper axis only, near $\nu=1$, and with the sample tilted to 63$^\circ$.  c)   $S/f_{pk}$ vs $\nu$, near $\nu=1$, where $S$ is integrated spectrum Re($\sigma_{xx}$ vs $f$.    d)  $S/f_{pk}$ vs $\nu$, near $\nu=1/3$.   Dashed lines  in c) and d) are  predicted $S/f_{pk}$ calculated from quasiparticle charge density as described in the text.   
 }
\end{figure}

Parameters of the resonances near $\nu=1$ and $1/3$ are summarized in Fig. 3.   Within the picture of the resonances as pinning modes, the common features of the the resonances in the two different $\nu$ ranges can be readily interpreted.
 The first common feature of the resonances is that  $f_{pk}$ increases as    $|\tilde{\nu}|$ (and hence   the quasiparticle charge density)    decreases, as shown for $\nu=1$ in panel a, and for $\nu=1/3$ in panel b.   
 This behavior has been observed \cite{lidensity} in the low $\nu$ Wigner crystal (whose density is equivalent to the total density $n$ of the sample) on tuning $n$  with a backgate.     Predicted in  weak pinning theories \cite{fl,Chitra,fertig,FoglerHuse}, the increase of $f_{pk}$  is due to the reduction in the carrier-carrier interaction (which results in  softening of the crystal)  that occurs as the intercarrier spacing  is increased. This reduction in the carrier-carrier interaction 
allows the carriers to  distribute themselves closer  to minima in the disorder potential, raising the pinning energy per carrier and thus $f_{pk}$.       

The integrated  conductivity  of  the resonances near $\nu=1$  and $1/3$ is in rough  agreement with predictions based on
on the quasiparticle charge density, $ ne  \tilde{\nu}/\nu$.  Within an oscillator model \cite{fl} of the pinning mode,  the integrated  Re$[\sigma_{xx}(f)]$ of a resonance is $S =ne  (| \tilde{\nu}|/\nu) \pi f_{pk}/2B$,   so $S/f_{pk}$  is nearly   proportional to $| \tilde{\nu}|$.     
  Figs. 3c and d  show $S/f_{pk}$ vs $\nu$ for the resonance near $\nu=1$ and 1/3, with $S$ obtained by integrating over the experimental frequency range,  up to 3 GHz.     For the 1/3 data,  the spectra merge into a decreasing background for larger $ \tilde{\nu}$, as seen in Fig. 2;   in Fig. 3d    we present data on $S/f_{pk}$ for $| \tilde{\nu}|$ up to the largest value for which we resolve a peak in the spectrum, about  $| \tilde{\nu}|=0.015$.  Dashed lines in the plots  show    $S/f_{pk}$ calculated from the oscillator model without   any variable parameter.    For both quantum Hall effects, the experimentally determined $S/f_{pk}$ vs $\nu$ roughly follow or lie below   the calculated  results; differences between the calculation and the data can be due 
  to missed amplitude in the integration of the spectra, and to existence of some  the quasiparticles not in the crystal:  for example, some carriers might be    strongly bound to disorder for small  $| \tilde{\nu}|$, or beginning to form liquid puddles at larger $| \tilde{\nu}|$.   For large $|\tilde{\nu}|$,   $S/f_{pk}$ vs $\nu$    flattens in the data taken around $\nu=1$ (Fig. 3c)   but continues to increase with $|\tilde{\nu}|$  around $\nu=1/3$ (Fig. 3d).  The reason for this difference is unclear; it may be that  liquid states to take up the carriers without contributing to the resonance  are better developed  near $\nu=1$.

The  $\nu$ range of the resonance around $\nu=1/3$ is    much narrower than the range of the resonance  around $\nu=1$,
but can be understood in terms of composite fermion theory \cite{jaincforig,jainbook}, supporting the interpretation of    1/3 FQH resonance as  an analogous  pinning mode of  FQHE quasiparticles.     CFs have their own effective filling factor,    
$\nu_{CF}=\nu/(1-2\nu)$, for CFs composed of two flux quanta plus an electron, and 
 the 1/3 FQHE is an IQHE of such CFs at $\nu_{CF}= 1$.     
A Wigner solid in the 1/3 FQHE can be construed as composed of CFs  within their $\nu_{CF}=1$ IQHE.
and as such is evidence for CF interaction.    FQHEs of interacting CFs have been observed \cite{panfqhcf} as well, and 
 CF Wigner solids have been considered theoretically \cite{narevich,goerbigreentrant,jaincfwc} for low electron Landau fillings.  
Changing $\nu = 1/3  \pm 0.015$    to CF filling results in  $\nu_{\rm{CF}}=1\pm 0.15$.  This is in  agreement with the  range of the pinning mode around $\nu=1$, which we observe out to 0.16 away from 
 $\nu=1$.   Furthermore, this  $\nu$  is reasonably close to theoretical predictions of   $\sim1/7$ \cite{lamgirvin,kunwc} for  the upper $\nu$ of a Wigner solid in the lowest Landau level, and  to experimental results in  low disorder n type samples, with  insulating behavior \cite{reentrant} and pinning modes \cite{yongab} seen for $\nu$ as high as about 0.218.


  Comparison of  peak frequencies of     the resonances  near $\nu=1$ and  near $\nu=1/3$ 
  strongly implies that the two resonances are of similar origin.
If  spin effects on the $\nu=1 $ resonance are suppressed, the $f_{pk}$   for the two ranges of $\nu$  agree for the same quasiparticle number density $ \tilde{n}=ne   \tilde{\nu}/\nu \tilde{e}$, where the quasiparticle charge $\tilde{e}=e$ or $e/3$.   The  resonances around $\nu=1$   
in samples  of  this density have been shown  \cite{Zhu} to be sensitive to tilting the sample in the magnetic field, which has been interpreted  \cite{Zhu} as due to suppression of  skyrmion formation \cite{schmellereisentein,cotefertig},
due to the larger Zeeman energy produced by the larger total magnetic field for a tilted sample near $\nu=1$.   In addition to $f_{pk}$ vs $\tilde{n}$ and $\nu$ near 1/3,   Fig. 3b  shows   $f_{pk}$ near  $\nu=1$ vs $ \tilde{n}$  on   the upper axis,  for   the  sample tilted  $63^{\rm{o}}$  from perpendicular to the magnetic field. That angle \cite{tiltnote}, according to ref. \cite{Zhu}, is sufficient to fully suppress the skyrmion effects.   
There is apparent agreement between  $f_{pk}$ vs $ \tilde{n}$  in the IQHE and FQHE.   We stress that while the agreement is much improved for the tilted-field data, there is  agreement to within a factor of two even without tilting.  

The agreement between $f_{pk}$ vs $ \tilde{n}$ in the IQHE and FQHE can be   interpreted within the weak pinning picture, if we take the solid near $\nu=1/3$ to be formed by $e/3$-charged carriers with Coulomb interaction. 
$f_{\rm{pk}}$ has been calculated \cite{Chitra} 
for   a   Coluomb-interacting solid subject to  weak, randomly distributed  disorder centers with potential strength $V_0$,  as 
$f_{\rm{pk}} \sim    (V_0^2/\mu \tilde{e}B\xi^6 )$, 
 where $\tilde{e}$ is the charge of a carrier, $\mu$ is the shear modulus of the solid, and $\xi$ is an effective disorder correlation length, which is the larger of the carrier wave function size  (approximately the magnetic length in the low $|\tilde{\nu}|$  limit) or the size of the disorder centers.    
The disorder, which has been proposed \cite{fertig} to be due to interface roughness,  is electrostatic, so $V_0$ must be proportional to $\tilde{e}$, and the classically calculated $\mu\sim (\tilde{n})^{3/2}  \tilde{e}^2$ \cite{bonsall}, so $V_0^2/\mu$ is $ \tilde{e}$-independent.    The factor $\tilde{e}B$ should also be  the same for the $\nu=1/3$ and $1$ effects.   The interpretation in terms of the weak pinning theory is then that the  effective disorder correlation length  $\xi$ is the same in the solids 
 around $\nu=1 $ and  $1/3$.

To summarize, for $\nu$ within a range of $\pm 0.015$ about $1/3$, we found that the real diagonal conductivity spectra of a 2DES exhibit microwave resonances. By analogy to similar resonances previously found near integer $\nu$, we interpret the resonances found near $\nu=1/3$ as the pinning modes of a disorder-pinned solid phase of  CFs,  or  quasiparticles of the 1/3 FQHE.    

We thank P. Littlewood, Kun Yang, and F. D. M. Haldane for helpful discussions, and we thank G. Jones, J. Park, T. Murphy and E. Palm for experimental assistance. This work was supported by DOE Grant Nos. DE-FG21-98-ER45683 at Princeton, DE-FG02-05-ER46212 at NHMFL. NHMFL is supported by NSF Cooperative Agreement No. DMR-0084173, the State of Florida and the DOE.


\begin{references}
\bibitem{tiltnote} At 63$^{\rm o}$, the $\nu$ range of the $\nu=1$   resonance and the $S/f_{pk}$ vs $\nu$ are essentially the same as  at $0^{\rm o}$.  Further, the 63$^{\rm o}$ $f_{pk}$ vs $\tilde{n}$ near $\nu=1$ agrees with that for the pinning mode near $\nu=2$. The $\nu\sim 2$ pinning mode in this sample is itself independent of tilt  from 0$^{\rm o}$ to 63$^{\rm o}$, as shown in ref 21.
\end{references}
\end{document}